\documentclass[twoside]{article}
\usepackage{fleqn,espcrc2}
\usepackage{graphicx}

\newcommand{\AmS}{{\protect\the\textfont2
   A\kern-.1667em\lower.5ex\hbox{M}\kern-.125emS}}
\usepackage{graphicx}%style pour inserer figures en eps
\usepackage{here}%pour inserer figures et tables a emplacement
\def\beq{\begin{equation}}
\def\eeq{\end{equation}}
\def\bea{\begin{eqnarray}}
\def\eea{\end{eqnarray}}
\def\bq{\begin{quote}}
\def\eq{\end{quote}}

\def\bear{\begin{array}}
\def\eear{\end{array}}
\def\nnb{\nonumber}
\def\ga{\left(}
\def\dr{\right)}

\def\rar{\rightarrow}
\def\Lrar{\Longrightarrow}

\def\nnb{\nonumber}
\def\la{\langle}
\def\ra{\rangle}
\def\nin{\noindent}
\def\ba{\begin{array}}
\def\ea{\end{array}}

\def\b{\bullet}

\def\mb{\overline{m}}

\def\gam5{\gamma_5}

\title{\bf{Light scalar mesons in QCD\thanks{Talk given at QCD 08 (7-12th july, Montpellier) and Scientific part of the invited talk for a tribute to Paco Yndurain 
at Confinement8 (3-7th september, Mainz) }
} 
}
\author{ Stephan Narison\thanks{E-mail:
snarison@yahoo.fr}\address{\footnotesize Laboratoire de Physique Th\'eorique et Astroparticules,\\
Universit\'e de Montpellier II
Place Eug\`ene Bataillon,
34095 - Montpellier Cedex 05, France.
}}
\begin{document}
\pagestyle{plain}
\begin{abstract}
\noindent
I present a mini-review of  the masses and couplings of the bare (unmixed) light scalar mesons : $\bar qq, ~(\overline{qq})(qq),~ (\bar qq)(\bar qq), ~gg$ from QCD spectral sum rules (QSSR) and low-energy theorems (LET) which we compare with recent lattice calculations when available. Some unbiased comments on  the different scenarios are given. The possiblity for the $\sigma (0.6)$ to be mostly a gluonium/glueball with a huge violation of the OZI rule in its decay is discussed. This review complements and updates the ones presented earlier \cite{SNG1}. Despite some progresses, the internal structure of the light scalar mesons remain puzzling, and some further efforts are required. It will be more fun at LHC if the Higgs of the Standard Model is a $\sigma$-like resonance.
\vspace*{2mm}
\noindent
\end{abstract}
\maketitle
\section{Introduction}
\nin
The nature of scalar mesons continues to be an intriguing problem in QCD.
Experimentally, around 1 GeV, there are well established scalar mesons with isospin $I=1$, the 
$a_0(980)$, and with isospin $I=0$, the $f_0(980)$ \cite{PDG}. Below 1 GeV,
the long-standing existence of  the wide $I=0~\sigma (600)$ is confirmed from $\pi\pi$ scattering \cite{LEUT,KAMINSKI} (see, however \cite{PETER}), while $K\pi$
scattering indicates the presence of the
$I=1/2~\kappa(840)$ \cite{PDG,PICH,BLACK2}, which is lower but larger than na\"\i vely expected from $SU(3)$ breaking expectations as a partner of the $a_0(980)$ in a $\bar qq$ scheme. 
The real quark and/or gluon contents of these scalar states are not fully understood which effective theories based on the (non) linear realization of chiral symmetry \cite{DIVECCHIA,LANIK,EFFECTIVE,ACHASOV,KYOTO}, 
may not help for clarifying this issue. In the following, we shall focus on the tests
of the $\bar qq,~(\bar qq)(\bar qq),~\overline{qq} qq$ and gluonium natures of these scalar mesons by confronting the recent experimental data with predictions on masses and couplings from QCD spectral sum rules (QSSR)
\cite{SVZ,SNB} complemented with some low-energy theorems (LET) \cite{SNG1,SNB,SNG,SNG0}, which we compare with lattice calculations \cite{LATTICE}--\cite{JAFFE2} and some other predictions when available. 
Previous reviews \cite{SNG1,MONTANET,KLEMPT,VENTO}) have been already dedicated to some of these studies. The present paper will
complement them and will update some recent developments in this field since 2006.
%\vspace*{-0.5cm}
%%%%%%%%%%%%%%%%%%%%%%%%%%%%%%%%%%%%%%%%
\section{The $I=1,~1/2$ scalar mesons}
%%%%%%%%%%%%%%%%%%%%%%%%%%%%%%%%%%%%%%%%%%%%%%%%%%%%
\nin
{\bf The \boldmath$a_0(980)$ and $\kappa(840)$ masses}\\
These channels are expected to be simpler as we do not expect to have any mixing with
a gluonium. If one assumes that these states are $\bar qq$ mesons, one can naturally
associate them to the divergence of the vector currents:
\bea
a_0(980)&\rar&\partial_\mu V^\mu_{\bar ud}\equiv (m_u-m_d):\bar u(i)d:~,\nnb\\
\kappa(840)&\rar&\partial_\mu V^\mu_{\bar us}\equiv (m_u-m_s):\bar u(i)s:~.
\label{current}
\eea 
Within the QSSR approach, the meson masses can be studied from the ratio of exponential Laplace/Borel sum rules \cite{SVZ,SNB}:
\beq\label{eq:sumrule}
{\cal R}_{n,n+1}=-{d\over d\tau}\log{\ga {\cal L}_n\equiv \int_0^\infty dt~t^n e^{-t\tau} {\rm Im}\psi_{\bar ud}(t)\dr}
\eeq
of the corresponding two-point correlator:
\beq
\psi_{\bar uq}(q^2)=i\int d^4x e^{iqx}\la 0|{\cal T}\partial_\mu V^\mu_{\bar
uq}(x)\partial_\mu V^\mu_{\bar uq}(0)^\dagger |0\ra~,
\eeq
where $\tau$ is the sum rule variable.
The observed {\it wrong} splitting between the  $a_0(980)$ and $\kappa(840)$ mesons can be 
understood from the crucial r\^ole of the four-quark condensates which reverses the splitting
as can be read from the approximate QSSR formula from ${\cal R}_{0,1}(\tau)$ \cite{SNG1}:
\bea
M^2_{\kappa}&\simeq& M^2_{a_0}+2\overline{m}_s^2- 8\pi^2m_s\la\bar ss\ra\tau_0\nnb\\
&+&{3\over 2} {1408\over 81}\pi^3\rho\alpha_s\ga \la \bar ss\ra^2-\la\bar uu\ra^2\dr\tau^2_0\nnb\\
&-&{1\over
3}M^2_\kappa\Gamma_\kappa^2\tau_0,
 \eea
where all different
parameters including the $a_0$ mass  are evaluated at the sum rule optimization scale $\tau_0\approx 1~ {\rm GeV}^{-2}$; $\rho\simeq
2\sim 3$ \cite{LAUNER} indicates the deviation from the vacuum saturation of the four-quark condensate; $\la \bar
ss\ra/\la\bar uu\ra\simeq 0.8$ measures the $SU(3)$ breaking of the quark condensate \cite{SNB}.  The last term is the finite width correction of the $\kappa$ which decreases its mass by about 20 MeV for $\Gamma_\kappa \approx 300$ MeV. 
 From the previous
analysis, one can deduce:
\beq
M_{a_0}\simeq 930~{\rm MeV}~~~~~~{\rm and}~~~~~~ M_{\kappa}\simeq 920~{\rm MeV}~,
\eeq
with about 10\% error, in good agreement with recent data \cite{PDG}.  A na\"\i ve  non-relativistic quark model gives analogous $a_0$ mass prediction but fails to reproduce the wrong splitting, which can be due to the unclear r\^ole of high-dimension operators  in this approach.
\\
Using similar QSSR analysis for the four-quark $(\overline{qq})(qq),~ (\bar qq)(\bar qq)$ operators for describing the scalar mesons, one obtains a value \cite{LATORRE,FA0,SNB}:
\beq
M_{4q}\approx 1~{\rm GeV}~,
\eeq
while a lattice calculation in a quenched approximation obtains analogous result for $M_{4q}$ \cite{JAFFE2}. \\
{\it   { \bf  --} QSSR can reproduce the wrong spiltting of the $a_0$ and $\kappa$ in te $\bar qq$ scheme which is not the case of the non-relativistic quark model. This result can be checked on unquenched lattices. At present, the alone evaluation of the spectrum cannot select the right quark content of these mesons.} \\
%%%%%%%%%%%%%%%%
{\bf $\b$ The decay constants}\\
The decay constant $f_{a_0}$ of the $a_0$ normalized as:
\beq
\la 0\vert \partial_\mu V^\mu_{\bar ud}\vert a_0\ra\equiv \sqrt{2} f_{a_0}M^2_{a_0}~,
\eeq
in the same way as $f_\pi=92.4$ MeV has been estimated several times in the literature.
The result from the exponential sum rule ${\cal L}_0(\tau)$ \cite{SNG1}:
\beq
f_{a_0}\simeq (1.6\pm 0.5)~{\rm MeV}~,
\eeq
leads to a value of the $u$-$d$ quark mass difference \cite{SCAL,SNL,SNB} consistent with recent lattice calculations.
Using $SU(3)$ symmetry and the almost degeneracy of the $a_0$ and $\kappa$ masses, one obtains with
a good accuracy:
\beq
{f_{\kappa}\over f_{a_0}}\simeq {m_s-m_u\over m_d-m_u}\simeq 40~,
\eeq
a ratio which is expected from the ChPT approach \cite{LEUT2}. \\
{\it 
{\bf --} 
We do not clearly see how the four quark  and diquark antidiquark schemes can reconcile with the explicit breaking of chiral symmetry by the light quark masses appearing in Eq. (\ref{current}). }\\
%%%%%%%%%%%%%%%%%%
{\bf $\b$ The hadronic couplings}\\
The $a_0$ and $\kappa$ hadronic couplings have been obtained using either a vertex sum rule
\cite{PAVER,SNA0} or/and $SU(3)$ symmetry rotation \cite{BN}. The leading order vertex sum rule results
are:
\bea
g_{a_0K^+K^-}&\simeq& {8\pi^2\over 3\sqrt{2}}{m_s\la\bar ss\ra\over M^2_Kf_K}\ga 1-{2\over r}\dr\simeq
3~{\rm GeV}~,\nnb\\
{g_{a_0K^+K^-}\over g_{\kappa K^+\pi^-}}&\simeq& e^{-(M_K^2-m_\pi^2)\tau_0}{\ga 1-{2\over r}\dr}
\simeq 1.17~,
\eea
where we have used \cite{SNB}: $m_s\la\bar ss\ra\simeq -0.8 M^2_Kf^2_K$, $r\equiv \la \bar ss\ra/\la\bar
uu\ra\simeq 0.8.$ and $\tau_0\simeq 1$ GeV$^{-2}$. We expect an accuracy of about 20\% (typical for the
3-point fonction sum rules) for these estimates. Using the
$SU(3)$ relation: 
\beq
g_{a_0\eta\pi}\simeq \sqrt{2\over 3}g_{a_0K^+K^-}~
\eeq
one obtains \footnote{Here and in the following, we use the normalization:
$\Gamma \ga S\rar PP'\dr \simeq {\vert g_{SPP'}\vert^2\over 16\pi M_{S}}\ga
1-{(M_{P}+M_{P'})^2\over M^2_{S}}\dr^{1/2}$, for a scalar meson $S$ decaying into  two pseudoscalar mesons $P$ and $P'$.}:
\bea
\Gamma \ga a_0\rar\eta\pi\dr \simeq 84~{\rm MeV}~, 
\eea
in agreement with the range of data from 50 to 100 MeV given by PDG \cite{PDG}.
Using the previous value of the $\kappa$ coupling, one can deduce:
\bea\label{eq: kappawidth}
\Gamma(\kappa\rar K\pi)\simeq {3\over 2}\Gamma(\kappa\rar K^+\pi^-)\simeq 104~{\rm MeV}~,
\eea
which is about a factor 4 smaller than the present data \cite{PDG}, but is a typical value for the width of a $\bar qq$ state.\\
The result for the hadronic coupling
$g_{a_0K^+K^-}$ in the four-quark scenario depends crucially on the operators describing the $a_0$ and can range
from 1.6 GeV
\cite{MARINA} to (5--8) GeV \cite{FA0}. However, the prediction for the $\eta\pi$ can agree with
the data \cite{FA0} depending on the size of the operator mixing parameter, while the four-quark prediction of the $\kappa$ hadronic width can lead to a half a value of the data. 
%Therefore, an eventual selection of the two approaches will be a precise measurement of $g_{a_0K^+K^-}$\\
\\
{\it 
%{\bf   \small Comments:} 
{\bf --} A strong deviation from the $SU(3)$ relation of the $a_0$ hadronic couplings does not favour the $\bar qq$ interpretation of the $a_0$, which seems not be the case \cite{MENES1}. One may question either  the validity of the $\bar qq$ or four-quark scheme for the $\kappa$ or a better understanding of its hadronic width from the data. This experimental question requires a clean separation of the direct coupling (resonance) of the $\kappa$ and the rescattering $K\pi$ term like is the case of the $\sigma$ meson \cite{MENES,MENES1} discussed later on.} \\
%%%%%%%%%%%%%%%%%%%%%%%%%%%%%%%%%
{\bf $\b$ The $\gamma\gamma$ width}\\
The $\gamma\gamma$ width of the $a_0$ has been evaluated using vertex sum rules within the $\bar qq$ 
and four-quark assignements of this meson, with the result \cite{FA0} \footnote{This approach leads to a successful predicition
of the well-known $\pi^0\to\gamma\gamma$ width.}:
\beq
\Gamma a_0(\bar qq)\rar \gamma\gamma\simeq (1.6\sim 2.6)~{\rm keV}~,
\eeq
and :
\beq
\Gamma a_0(4q)\rar \gamma\gamma\simeq (2\sim 5)\times 10^{-4}~{\rm keV}~,
\eeq
where the size of the ratio is of the order of $(\alpha_s/\pi)^2$, indicating that the four-quark assignement prediction is small (see also \cite{ACHASOV}).\\
{\it {\bf --} None of the two schemes can explain the data of ($0.24\pm 0.08$) keV  \cite{PDG}, which is not the case of an effective 
approach based on a kaon hadronic tadpole mechanism for $SU(2)$ breakings \cite{BN} or on a similar
model involving kaon loops \cite{ACHASOV}. Unfortunately, a connection between these approaches and the $\bar qq$ or $4q$ scheme is not fully understood. A separate measure of  the direct coupling and of the $\bar KK$ rescattering terms may clarify this issue.}

%%%%%%%%%%%%%%%%%%%%%%%%%%%%%
\section{The $I=0$ bare scalar mesons}
\nin
The isoscalar scalar states are especially interesting in the framework of QCD since, in this anomalous $U(1)_V$
channel, their interpolating  operator is the trace of the energy-momentum tensor:
\beq
\theta_\mu^\mu=\frac{1}{4}\beta(\alpha_s) G^2+\sum_{i\equiv u,d,s} [1+\gamma_m(\alpha_s)]
m_i\bar\psi_i\psi_i~,
\label{eq:thetamumu}
\eeq
where $G^a_{\mu\nu}$ is the gluon field strengths, $\psi_i$ is the 
quark field; $\beta(\alpha_s)\equiv\beta_1\ga \alpha_s/\pi\dr+...$ and
$\gamma_m(\alpha_s)\equiv\gamma_1\ga \alpha_s/\pi\dr+...$ are respectively the QCD $\beta$-function and 
quark mass-anomalous dimension ($\beta_1=-1/2(11-2n/3)$ and $\gamma_1=2$ for $n$ flavours). In the chiral
limit
$m_i=0$,
$\theta_\mu^\mu$ is dominated by its gluon component $\theta_g$, like 
is the case of the $\eta'$ for the
$U(1)_A$ axial-anomaly, explaining why the $\eta'$-mass does not 
vanish for $m_i=0$ \cite{WITTEN}, though it loves to couple to ordinary mesons. Then, it is natural to expect that these $I=0$ scalar states 
are glueballs/gluonia or have at least a strong
glue component in their wave function. This gluonic part of $\theta^\mu_\mu$ can be identified with the $U(1)_V$ term \cite{DIVECCHIA} of the effective lagrangian based on a $U(3)_L\times U(3)_R$ linear realization of chiral symmetry (see e.g.
\cite{LANIK,EFFECTIVE,ACHASOV,KYOTO}). \\
%%%%%%%%%%%%%%%%%%%%%%%%%%%%%%%%%%%%%%%%%%%%%%%%%%%%%%
{\bf $\b$ Unmixed \boldmath$I=0$ scalar $\bar qq$ mesons}\\
%%%%%%%%%%%%%%%%%%%%%%%%%%%%%%%%
We shall be concerned with the mesons $S_2$ and $S_3$ mesons associated respectively to the quark
currents:
\beq 
J_2=m:{1\over \sqrt{2}}(\bar uu+\bar dd):~~~{\rm and}~~~J_3=m_s:\bar ss:~. 
\eeq
From the good realization of the $SU(2)$ flavour symmetry ($m_u=m_d$ and $\la\bar uu\ra=\la\bar dd\ra)$, 
one expects a degeneracy between the $a_0$ and $S_2$ states:
\beq
M_{S_2}\simeq M_{a_0}\simeq 930~{\rm MeV}~,
\eeq
while its hadronic coupling is \cite{BN,SNG}:
\beq\label{eq: s2pipi}
g_{S_2\pi^+\pi^-}\simeq {16\pi^3\over 3\sqrt{3}}\la\bar uu\ra \tau_0 e^{M^2_2\tau_0/2}\simeq 2.46~{\rm GeV}~.
\eeq
leading to:
%corresponding to \footnote{We use
%the normalization:
%$$
%\Gamma(\sigma_B\rar\pi\pi)={3\over 2}\frac{|g_{\sigma_B\pi^+\pi^-}|^2}
%{16\pi M_{\sigma_B}}\ga{1-\frac{4m^2_\pi}{M^2_{\sigma_B}}}\dr^{1/2}.
%$$.}:
\beq
\Gamma (S_2\rar \pi^+\pi^-)\simeq 120~{\rm MeV},
%\nnb\\&&\Gamma (S_2\rar \gamma\gamma)\simeq (0.67\pm 0.04)~{\rm keV}~.
\eeq
Using $SU(3)$ symmetry, one can also deduce:
\beq\label{eq: s2kk}
g_{S_2K^+K^-}\simeq {1\over 2}g_{S_2\pi^+\pi^-}\simeq 1.23~{\rm GeV}~.
\eeq
%The $S_2$ $\gamma\gamma$ width can be deduced from the one of the $a_0(\bar qq)$ obtained previously, through the
%non-relativistic relation (ratio of the square of quark charges):
%\beq\label{eq:gamma}
%\Gamma_{S_2\rar\gamma\gamma}\simeq {25 \over 9} \Gamma_{a_0\rar\gamma\gamma}\simeq (0.7\pm 0.2)~{\rm keV}~.
%\eeq
The mass of the mesons
containing a strange quark is predicted to be \cite{SNG}:
\beq
M_{S_3}/M_{\kappa}\simeq 1.03\pm 0.02~~~~\Lrar M_{S_3}\simeq 948~{\rm MeV}~,
\eeq
if one uses $M_{\kappa}=920$ MeV \footnote{In \cite{SNG} a higher value has been obtained because one has
used as input the experimental mass $K^*_0=1430$ MeV.}, while its coupling to $K^+K^-$
% and to $\gamma\gamma$ 
is 
\cite{SNG}:
\beq\label{eq: s3kk}
g_{S_3K^+K^-}\simeq (2.7\pm 0.5)~{\rm GeV}~.
%\nnb\\&&\Gamma (S_3\rar \gamma\gamma)\simeq (0.40\pm 0.04)~{\rm keV}.
\eeq
{\it {\bf --} The predictions for the hadronic widths suggest that the na\"\i ve $\bar qq$ assignement of the $\sigma(600)\equiv S_2$ does not fit the data.}
\\
%%%%%%%%%%%%%%%%%%%%%%%%%%%%
{ \bf $\b$ Gluonia masses from QSSR} \\
%%%%%%%%%%%%%%%%%%%%%%%%%%%%
 \nin
Masses of the bare unmixed scalar gluonium can be determined from the two Laplace unsubtracted (USR) and subtracted (SSR) sum rules \cite{VENEZIA}:
\bea
{\cal L}_0(\tau)&=&{1\over\pi}\int_0^\infty dt e^{-t\tau}{\rm Im} \psi(t)~,\nnb\\
{\cal L}_{-1}(\tau)&=&-\psi(0)+{1\over\pi}\int_0^\infty {dt\over t} e^{-t\tau} {\rm Im} \psi(t)~,
\eea
%\vspace*{0.2cm}
of the two-point correlator $\psi(q^2)$ associated to $\theta^\mu_\mu$ defined in Eq. (\ref{eq:thetamumu}).
The subtraction constant $\psi(0)=-16(\beta_1/\pi )\la \alpha_s G^2\ra$ expressed in terms of the gluon condensate \cite{NSVZ} $\la \alpha_s G^2\ra= (0.07\pm 0.01)~{\rm GeV}^4$ \cite{SNG2,YND,SNB} affects strongly the USR analysis which has lead to apparent discrepancies in the previous literature when a single resonance is introduced into the spectral function \cite{SNG1}. The SSR being sensitive to the high-energy region $(\tau\simeq 0.3$ GeV$^{-2}$) predicts  \cite{SNG}:
\beq
M_G\simeq (1.5\sim 1.6)~{\rm GeV}~,
\eeq
comparable with the quenched lattice value \cite{QUENCHED}, while the USR stabilizes at lower energy $(\tau\simeq 0.8$ GeV$^{-2}$) and predicts a low-mass gluonium \cite{SNG0} \footnote{Notice that using a similar USR, the trigluonium $0^{++}$ mass associated to the scalar operator $g f_{abc}G^aG^bG^c$
is found to be about 3.1 GeV \cite{PABAN,SNB} and has a tiny mixing with the digluonium.}:
\beq
M_{\sigma_B}\simeq (0.9\sim 1.1) ~{\rm GeV}~,
\eeq
comparable with the unquenched lattice value \cite{LATTICE} but higher than the one \cite{FRASCA}
using a strong coupling calculation of the gluon propagator \footnote{A more direct comparison requires the evaluation of the gluonic two-point  in this approach.}.
Furthermore, the consistency
of the USR and SSR can be achieved by a two-resonances ($G$ and
$\sigma_B$) + ``QCD continuum" parametrization of the spectral function
\cite{VENEZIA}\footnote{In \cite{SNG} the QCD continuum has also been
modelized by a $\sigma'_B$ (radial excitation of the $\sigma_B$), which
enables to fix the decay constant $f_{\sigma_B,\sigma'_B}$ once the $\sigma_B,~\sigma'_B$
masses are introduced as input.}. \\
{\bf --} {\it A recent QSSR analysis of the same
gluonium correlator  including direct instantons and using  a two-resonance parametrization \cite{STEELE} confirms the previous mass values, while the one using a single resonance  \cite{FORKEL,VENTO} gives the mean value of the two masses.The small effects of the direct instanton in the mass predictions is expected from the smallness of the extra $1/q^2$ term induced by a tachyonic gluon mass which mimics instanton effects in this approach \cite{ZAK},
and which is necessary for solving the sum rule scale hierarchy between the gluonia
and of the usual $\rho$ meson \cite{CNZ}.}\\
%%%%%%%%%%%%%%%%%%%%%%%%
{\bf $\b$ OZI violation in $\sigma_B \to\pi\pi$}\\
%%%%%%%%%%%%%%%%%%%%%%%%
 \nin
The $\sigma_B\pi\pi$ coupling can be obtained from the vertex function:
\beq
V[q^2\equiv (q_1-q_2)^2]\equiv \la \pi_1\vert \theta_\mu^\mu\vert \pi_2\ra~,
 \eeq
 obeying a once subtracted dispersion relation:
 \beq
 V(q^2)=V(0)+q^2\int_{4m_\pi^2}^\infty {dt \over t}{1\over t-q^2-i\epsilon}{1\over \pi}{\rm Im} V(t)~.
 \eeq
with the condition:  $V(0)={\cal O}(m^2_\pi) \rar 0$ in the chiral limit. Using also the fact that $V'(0)=1$,  one can then derive the two sum rules:
\beq
\sum_{S=\sigma_B,...} g_{S\pi\pi}\sqrt{2}f_S=0~,~
{1\over 4}\sum_{S=\sigma_B,...} g_{S\pi\pi}{\sqrt{2}f_S\over M_S^2}=1~
\label{eq: sigmapipi}
\eeq
where $f_S$ is the decay constant analogue to $f_\pi$. The 1st sum rule requires the existence of at least two resonances coupled strongly to $\pi\pi$. Considering the $\sigma_B$ and $\sigma'_B$ but neglecting the small $G$-coupling to $\pi\pi$
as indicated by GAMS \cite{GAMS}, one predicts in the chiral limit \cite{VENEZIA,SNG} :
\beq
  |g_{\sigma_B\pi^+\pi^-}|\simeq 
    |g_{\sigma_B K^+K^-}| \simeq (4\sim 5)~{\rm GeV}~,
    \label{pipicoupling}
    \eeq
a universal coupling, which will imply a large width~\footnote {However, the analysis in Ref. \cite{SNG} also indicates  that $\sigma_B$ having a mass below 750 MeV cannot be wide ($\leq 200$ MeV) (see also some of Ref. \cite{SNG1})
due to the sensitivity of the coupling to $M_\sigma$. Wide low-mass gluonium has also been obtained
using QSSR (1st ref. in \cite{SNG1}), an effective Lagrangian \cite{LANIK} and LET \cite{NSVZ}.}:
%\vspace*{-0.35cm}
\beq
\Gamma_{\sigma_B\to\pi^+\pi^-} 
%\equiv  {|g_{\sigma_B
%  \pi^+\pi^-}|^2\over {16\pi M_{\sigma_B}}}\ga 1-{4m^2_\pi \over
% M^2_{\sigma_B}}\dr^{1/2}
 \simeq0.7~ {\rm GeV}~.
  \label{eq:scalarwidth} 
\eeq
{\bf --} {\it The large width into $\pi\pi$ is a typical OZI-violation  due to non-perturbative effects expected to be important in the region below 1 GeV, where perturbative arguments valid in the region of the G(1.5) 
cannot be applied. This result can be tested using lattice unquenched calculations of the width.}\\
%%%%%%%%%%%%%%%%%%%%%%%%%%%%%%%%%%
{\bf $\b$ \boldmath $G(1.5)$ widths}\\
We shall not discuss the derivation of these widths here as they are done in details in \cite{VENEZIA,SNB,SNG1}. An analogous low-energy theorem \cite{VENEZIA} like the one in Eq. (\ref{eq: sigmapipi}) leads to strong   couplings of the $G(1.5)$ in the $U(1)_A$ channels $\eta'\eta'$ and $\eta\eta'$, while, one expects its weaker couplings to $\pi\pi$ contrary to $\sigma_B$, which almost 
saturates the vertex sum rule in Eq. (\ref{eq: sigmapipi}). \footnote{These features are expected for a {\it pure gluonium} where a perturbative argument like the chiral coupling to $\pi\pi$ and $\bar KK$ can also hold \cite{CHANOWITZ}.}. Characteristic glueball decay widths in these
channels are \cite{VENEZIA,GERS}:
\beq
\Gamma(G\rar\eta'\eta)\simeq (5\sim 10)~{\rm MeV}~, ~~~\frac{\Gamma_{G\eta\eta}}{\Gamma_{G\eta\eta'}}\simeq 0.22~.
\eeq
%%%%%%%%%%%%%%%%%%%%%%%%%%%%%%%%%%%%%%%%%%%%%%%%%%%%
In this approach, where the $\sigma_B$ is the lowest mass gluonium, one also expects that the G(1.5) decays into $4\pi$ via $\sigma_B$ pairs with:
\beq
|g_{G\sigma\sigma}|\approx 1.3\sim 3.7~\mbox{GeV}\Lrar \Gamma_{G\to4\pi}\approx 7-55~{\rm MeV}
\eeq
%%%%%%%%%%%%%%%%%%%%%%%%%%%%
%\vspace*{-0.3cm}
{\bf $\b$ Gluonia couplings to $\gamma\gamma$}\\
%%%%%%%%%%%%%%%%%%%%%%%%%%%%
 \nin
These couplings can be derived by identifying the Euler-Heisenberg effective Lagrangian for  $gg\to \gamma\gamma$ via a quark
constituent  loop to the scalar one: $
{\cal L}_{S\gamma\gamma}= g_{S\gamma\gamma}SF_{\mu\nu}^{(1)}F_{\mu\nu}^{(2)},
$
 which leads to the sum rule \footnote{This sum rule has been used in \cite{NSVZ2} for the charm quark.}:
\beq
g_{S\gamma\gamma} \simeq {\alpha\over 60}\sqrt{2}f_S M_S^2\ga{\pi\over-\beta_1}\dr \sum_{q=u,d,s}Q^2_q/M^4_q~, 
\eeq
where $Q_q$ is the quark charge in units of $e$; $M_{u,d}\approx M_\rho/2$ and   $M_{s}\approx M_\phi/2$;  are constituent masses; $S$ refers to gluonium ($\sigma_B,...)$. Then, one predicts the couplings:
\beq
g_{\sigma_B\gamma\gamma}\approx g_{\sigma'_B\gamma\gamma}\simeq g_{G\gamma\gamma} \simeq (0.4\sim 0.7)\alpha ~{\rm GeV}^{-1}~,
\eeq
and the corresponding widths \footnote{Due to their $M^3$-dependence, the  widths of the $\sigma'_B$ and $G$ can be much larger: $(0.4\sim 2)$ keV. These widths induce a tiny effect of $(3-9) \times 10^{-11}$ to the muon $(g-2)$ \cite{SNANOM} and cannot be excluded.} :
\beq
\Gamma_{\sigma_B\to\gamma\gamma}\equiv  {|g_{\sigma_B
\gamma\gamma}|^2\over 16\pi }M^3_{\sigma_B}\simeq (0.2\sim 0.6)~{\rm keV}~.
\eeq
Alternatively, one can match the $k^2$ dependence of the two sides of $ \la 0| \theta_\mu^\mu|\gamma_1\gamma_2 \ra $ in order to derive the sum rule \cite{CHANO,LANIK}:
% For a self-consistency check of the previous results, one can introduce their values into the sum rule:
  \beq
 {1\over 4}\sum_{S=\sigma_B,...} g_{S\gamma\gamma}\sqrt{2}f_S={\alpha R\over 3\pi}~,
 \eeq
 which one can use for checking the a self-consistency of the previous results ($R\equiv 3\sum Q^2_q$) \cite{VENEZIA}.\\
  %%%%%%%%%%%%%%%%%%%%%%%%%%%%%%%%%%%%%
%  \vspace*{-0.1cm}
{\bf $\b$ QCD tests of the $\sigma /f_0(600)$= a gluonium ?} \\
%%%%%%%%%%%%%%%%%%%%%%%%%%%%%%%%%%%%%%
 \nin
 The first question which comes in mind is: {\it  how can one compare the theoretical results from QSSR and LET obtained in the real axis
 with the measured value of the complex pole parameters  from $\pi\pi,~\pi K$ and $\gamma\gamma$ scatterings ?} Some approximate answers to this question are given in the literature, either by using a Breit-Wigner parametrization of the data \cite{MENES}, or by using \cite{MENES} the on-shell mass obtained by imposing that the amplitude is purely imaginary at the phase 90$^0$ \cite{SIRLIN} \footnote{Unfortunately this procedure is expected to be only accurate for a narrow resonance.} or by arguing that the mass obtained from QSSR appears in the tree level amplitude of $\pi\pi$ scattering and becomes much lower in the unitarized amplitude \cite{LEUT,ACHASOV2}. In all cases, these results indicate that the 1 GeV mass from QSSR  can translate into the observed wide $\sigma(600)$. Using the complex pole mass, one can e.g. deduce the on-shell mass:
 \beq
 M_\sigma^{\rm os}\approx 0.92~{\rm GeV}~~~~{\rm and} ~~~~\Gamma_\sigma^{\rm os}\approx 1~{\rm GeV},
 \eeq
 in remarkable agreement with the previous QSSR and LET predictions. 
For (properly) comparing the $\gamma\gamma$ width of $\sigma_B$ with the one of $\sigma(600)$, one
can use the Mennessier model \cite{MENES,MENES1} for separating the direct resonance coupling from the $\pi\pi$ rescattering terms. Then, we can deduce the ``partial" $\gamma\gamma$ widths at the complex pole :
\bea
\Gamma_{\sigma\to\gamma\gamma}^{\rm dir}&\simeq& (0.13\pm 0.05)~{\rm keV}~,\nnb\\
\Gamma_{\sigma\to\gamma\gamma}^{\rm resc}&\simeq& (2.7\pm 0.4)~{\rm keV}~,
\eea
 and the total $\gamma\gamma$ width (direct + rescattering):
 \vspace*{0.3cm}
 \beq
 \Gamma_{\sigma\to\gamma\gamma}^{\rm tot}\simeq (3.9\pm 0.6)~{\rm keV}~,
 \label{radtot}
\eeq
These values are in the range of the ones obtained in the literature \cite{PENN1,OLLER,PRADES} but a clean comparison
is difficult to do as the separation between the direct and the rescattering terms is not explicit there. An (almost) similar model  \cite{ACHASOV07} leads to a result 2 times smaller for the total contributions than the one obtained here and a negligible direct $\gamma\gamma$ width of about 3.4 eV. A meaningful comparison can be done for the total contribution but requires the quotation of errors in the result of \cite{ACHASOV07}.  \\
In \cite{LAYSSAC} a clean separation of the two contributions is proposed by measuring the $C$-odd asymmetry in $e^+e^-\to \pi^+\pi^-$, which can be feasible at KLOE \cite{KLOE}. Improvements of the prevoius estimates need more precise data below 700 MeV, and an extension of the analysis to higher energies. Translating the previous results to the on-shell mass, one obtains:
\beq
\Gamma_{\sigma\to \gamma\gamma}^{\rm os,dir}
\approx (1.0\pm 0.4)~{\rm keV}~,
\label{eq:osgamwidth}
\eeq
in good agreement with the QSSR predictions for $\sigma_B$. \\
{\bf --} {\it  
The previous  ``overall agreement" without any free mixing parameter can favour a large gluon component in the $\sigma/f_0(600)$ wave function, which is naturally expected due to the $U(1)_V$ anomaly.
 Its large width into $\pi\pi$, indicates a strong violation  of the OZI rule,
 signals large non-perturbative effects in its treatment, and disfavours its $\bar qq$ interpretation. The latter predicts a narrower hadronic width and a much larger $\gamma\gamma$ width of about 5 keV which
 can be deduced from the one of the $a_0$.  In the same way, QSSR also  predicts, for a four-quark state, having the same mass of 1 GeV \cite{LATORRE,SNA0}, a tiny $\gamma\gamma$ width of about 0.4 eV \cite{SNA0}. Both $\bar qq$ and 4-quark scenarios are
disfavoured by the value of the direct coupling fitted from $\gamma\gamma$ scattering, while an explanation of the $I=0$ scalar channel without the inclusion of a gluonium/glueball associated to the $U(1)_V$-anomaly, appears to be unlikely. Another point which can disfavour the $4q$ scenario is the expectation of the weak coupling of the $\sigma$ to $\bar KK$ \cite{ACHASOV}, which seems not be the case from the analysis of \cite{MENES1,KAMINSKI2} where the
$\pi\pi$ and $KK$ couplings are almost equal as expected  from the LET results in Eq. (\ref{pipicoupling}).}
%%%%%%%%%%%%%%%%%%%
\section{Higher scalar meson masses}
%%%%%%%%%%%%%%%%%%%
\nin
Understanding the dynamics of the spectrum of the radial excitations within QCD is still unanswered and is the subject of hot activities. For the $I=0$ scalar mesons, the $f_0(1.37)$ is a good candidate for being the radial excitation of the $\sigma(600)$ where its hadronic width into $\pi\pi$ is found to be about $(300\sim 500)$ MeV \cite{VENEZIA,SNG,SNG1} and its $\gamma\gamma$ width of the order of $(10\sim 30)$ eV. It can be viewed as a high mass tail of the gluonic sigma ({\it red dragon}) in \cite{MINK},
which starts from the $\sigma(600)$ and continues to higher tail of about 1800 MeV. However, the experimental situation for the $f_0$(1.37) is not yet settled \cite{ANISO,OCHS4}.
%%%%%%%%%%%%%%%%%
\section{Meson-gluonium mixing}
%%%%%%%%%%%%%%%%%
\nin
From the present unquenched lattice result where the glueball mass has shifted down to 1 GeV, existing 
meson-gluonium schemes \cite{CLOSE} need to be revised. Hopefully, we have always considered \cite{SNG0,VENEZIA,SNG,SNG1} that
the lowest gluonium mass is below 1 GeV making still valid the meson-gluonium mixing discussed in \cite{BN,SNG1,SNB} where a maximal two-component mixing between the $S_2 (\bar qq)$ and $\sigma_B(gg)$ have been proposed for explaining the wide $\sigma(600)$ and narrow $f_0(980)$. Above 1 GeV, a three-component mixing \`a la CKM between the $G(1.5), ~\sigma'_B(1.37)$ and the radial excitation $S'_3(1.47)$ of the $\bar ss$ state have been proposed for explaining the $f_0(1.37), ~f_0(1.5)$ and $f_0(1.7)$, which can be updated.
%%%%%%%%%%%%%%%%
\section{Experimental tests}
%%%%%%%%%%%%%%%
\nin
In addition to the new one \cite{LAYSSAC} for measuring the $\pi\pi$ rescattering term in $e^+e^-\to \pi^+\pi^-$ discussed here and the reanalysis of $\gamma\gamma$ scattering experiments, some characteristic
tests for detecting glueballs can be found in \cite{SNG1} and some other existing reviews.
% \cite{SNG1,KLEMPT,VENTO}.
\section{General remarks}
\nin
The scalar sector of QCD remains complex and fascinating, while many problems remain unanswered after a half century. It will be more fun if the Higgs of the Standard Model is a $\sigma$-like resonance.
\section*{Acknowledgments}
%%%%%%%%%%%%%%%%
\nin
It is a pleasure to thank all my previous collaborators who have participated to some parts of the works reported in this review and Wofgang Ochs for reading the manuscript.
%%%%%%%%%%%%%%%%%%%%%%%%%%%%%%%%%%%

\end{document}